\documentclass[amssymb,aip,jap,numerical,superscriptaddress]{revtex4-1}

\usepackage{amsmath}
\usepackage{graphicx}
\usepackage{subfigure}
\usepackage{hyperref}

\newcommand{\lef}{\left(}
\newcommand{\rig}{\right)}
\newcommand{\deriv}{\textnormal{d}}

\begin{document}

\title{Resistivity scaling and electron relaxation times in metallic nanowires}
\author{Kristof Moors}
\email[Electronic mail: ]{kristof@itf.fys.kuleuven.be}
\affiliation{KU Leuven, Instituut voor Theoretische Fysica, Celestijnenlaan 200D, B-3001 Leuven, Belgium}
\affiliation{Imec, Kapeldreef 75, B-3001 Leuven, Belgium}
\author{Bart Sor\'ee}
\affiliation{Imec, Kapeldreef 75, B-3001 Leuven, Belgium}
\affiliation{Universiteit Antwerpen, Physics Department, Groenenborgerlaan 171, B-2020 Antwerpen, Belgium}
\author{Zsolt T\H okei}
\affiliation{Imec, Kapeldreef 75, B-3001 Leuven, Belgium}
\author{Wim Magnus}
\affiliation{Imec, Kapeldreef 75, B-3001 Leuven, Belgium}
\affiliation{Universiteit Antwerpen, Physics Department, Groenenborgerlaan 171, B-2020 Antwerpen, Belgium}

\date{\today}

\begin{abstract}
We study the resistivity scaling in nanometer-sized metallic wires due to surface roughness and grain-boundaries, currently the main cause of electron scattering in nanoscaled interconnects. The resistivity has been obtained with the Boltzmann transport equation, adopting the relaxation time approximation (RTA) of the distribution function and the effective mass approximation for the conducting electrons. The relaxation times are calculated exactly, using Fermi's golden rule, resulting in a correct relaxation time for every sub-band state contributing to the transport. In general, the relaxation time strongly depends on the sub-band state, something that remained unclear with the methods of previous work. The resistivity scaling is obtained for different roughness and grain-boundary properties, showing large differences in scaling behavior and relaxation times. Our model clearly indicates that the resistivity is dominated by grain-boundary scattering, easily surpassing the surface roughness contribution by a factor of 10.
\end{abstract}

\maketitle

\section{Introduction}
The resistivity scaling of metallic nanowires is very important for the application as interconnects in chips. An increase of resistivity causes many problems, e.g. increased heating, power consumption, signal delay and errors. One has realized already for quite some time that interconnect resistivity scaling is one of the major issues in further down-scaling of micro-electronic devices.\cite{ho2001future,steinlesberger2002electrical} Metallic nanowires with smaller diameters suffer from an increase in resistivity because of enhanced electron collisions. Two factors in small-diameter nanowires are generally assumed to be causing this increase of collisions: surface roughness of the wire boundary and grain-boundaries. The corresponding scattering processes are already known for a long time and many experiments have confirmed the importance of their contributions to the overall resistivity.

Intuitively, it is clear that the time between subsequent surface scattering events goes down for smaller diameters because an increasing surface to volume ratio. Similarly, time between subsequent grain-boundary collisions decreases, because there is a general trend of increasing grain-boundary density for nanowires of smaller cross-sections. Both surface and grain-boundary scattering events lead to a substantial loss of forward momentum of the conducting electrons, thereby increasing the resistivity. Using the classical Drude model, we can relate the time between subsequent scattering events $\tau$ to the resistivity as $\rho \propto \tau^{-1}$. This result is only an approximation as all the quantum-mechanical effects have been neglected.

Two standard models describe surface roughness and grain-boundary scattering more rigorously: the Fuchs-Sondheimer\cite{fuchs1938conductivity,sondheimer1952mean}(FS) and the Mayadas-Shatzkes\cite{mayadas1970electrical}(MS) model, respectively dealing with surface roughness and grain-boundary scattering (including surface roughness). The FS and MS models confirm the increase of resistivity for smaller diameters while predicting that the relaxation time is inversely proportional to the diameter of the interconnect $D$ (which refers to the smallest width or height in case of a rectangular cross-section), $\rho \propto D^{-1}$, as observed experimentally.\cite{steinhogl2002size,guillaumond2003analysis,steinhogl2004comprehensive,chawla2011electron} There is no consensus about the relative contributions of surface roughness and the grain-boundaries to the resistivity\cite{durkan2000size,wu2004influence,zhang2007analysis,graham2010resistivity} (see Josell\cite{josell2009size} for an overview of contributions), which is understandable because their relative importance depends on the properties of the surface and grain-boundaries, which, in turn, are very sensitive to the experimental set-up. Since one of both resistivity contributions cannot be excluded in general, it is very important to understand both surface roughness and grain-boundary scattering in interconnects in order to reduce it.

Although FS and MS resistivity scaling shows good agreement with current experimental data, there are two main reasons to doubt their validity for diameters below 10 nm, a regime in which experimental data is currently unavailable. First, one needs to rely on a free parameter $p$, known as the Fuchs parameter, describing the surface roughness scattering, in order to fit the models to experimental data. The Fuchs parameter is the probability for specular scattering at the boundary, whereas the electrons scatter diffusively at the boundary with a probability $1-p$. This parameter is often estimated to be 50$\%$ but a general way of calculating its value from the material and roughness properties is not known, especially below 10 nm diameters. The argument leading to the introduction of the Fuchs parameter is purely classical and neglects all the quantum mechanics that governs ultra-thin nanowires. Secondly, the FS and MS models invoke a continuum approximation for all the states in the calculation. The basis states in the system are labeled continuously in every direction of the material, resembling the bulk. This is definitely not valid for nanowires with very small diameters, for which quantized sub-bands must be considered.

We propose a model that solves the above issues, using the Boltzmann transport equation and an effective mass approximation. The scattering mechanism due to surface roughness is based on Ando's model\cite{ando1982electronic} avoiding the Fuchs parameter whose relation to the microscopic wire properties is unclear. Ando's model eliminates this parameter by using quantum-mechanical perturbation theory to describe surface roughness scattering. Being able to provide transport properties as a function of surface roughness characteristics like the standard deviation and the correlation length of the boundary deformation, it was originally devised to describe a 2D electron gas, but more recently it has also been used to study other systems, such as semiconductor nanowires and MOSFET's.\cite{esseni2004modeling,mazzoni1999surface,jin2007modeling,jin2007modeling2} According to the paper of Mayadas and Shatzkes, grain boundaries are modeled in the most simple way as surfaces of potential energy perpendicular to the transport direction.\cite{mayadas1970electrical}

We apply our model to rough square, metallic nanowires containing grain-boundaries and look specifically at the contribution to the resistivity and corresponding scaling behavior in the few nanometer regime, for which we are able to retrieve the relaxation times for each sub-band state correctly.

\section{Boltzmann transport equation - relaxation time approximation}
\label{sectionBTERTA}

\setlength{\unitlength}{1.2cm}
\begin{figure}[htb]
\subfigure[\ Grain boundaries]{
\begin{picture}(7.0,4.25)(-3.5,-0.25)
\linethickness{0.25mm}
\put(-2.5,0.5){\line(1,0){6}}
\put(-2.5,2.5){\line(1,0){6}}
\put(-3,1.0){\line(1,0){6}}
\put(-3,3.0){\line(1,0){6}}
\put(-2.5,0.5){\line(-1,1){0.5}}
\put(-2.5,2.5){\line(-1,1){0.5}}
\put(-2.6,0.4){\vector(-1,1){0.5}}
\put(-3.1,0.9){\vector(1,-1){0.5}}
\put(-2.5,0.5){\line(0,1){2}}
\put(-3.15,1.0){\vector(0,1){2}}
\put(-3.15,3.0){\vector(0,-1){2}}
\put(-3,1.0){\line(0,1){2}}
\put(3.5,0.5){\line(-1,1){0.5}}
\put(3.5,2.5){\line(-1,1){0.5}}
\put(3.5,0.5){\line(0,1){2}}
\put(3,1.0){\line(0,1){2}}
\put(-3.1,0.55){\makebox(0,0){\small $D$}}
\put(-3.4,1.9){\makebox(0,0){\small $D$}}
\put(0.5,0.2){\makebox(0,0){\small $L_z$}}
\put(-2.5,0.4){\vector(1,0){6}}
\put(3.5,0.4){\vector(-1,0){6}}
\put(0.1,2.2){\makebox(0,0){\small $E_z, J_z$}}
\put(-0.5,1.9){\vector(1,0){1.2}}
\put(-1.5,0.5){\line(-1,1){0.5}}
\put(-1.5,1.0){\line(-1,1){0.5}}
\put(-1.5,1.5){\line(-1,1){0.5}}
\put(-1.5,2.0){\line(-1,1){0.5}}
\put(-1.5,2.5){\line(-1,1){0.5}}
\put(-1.5,0.5){\line(0,1){2}}
\put(-1.625,0.625){\line(0,1){2}}
\put(-1.75,0.75){\line(0,1){2}}
\put(-1.875,0.875){\line(0,1){2}}
\put(-2,1.0){\line(0,1){2}}
\put(-0.7,0.5){\line(-1,1){0.5}}
\put(-0.7,1.0){\line(-1,1){0.5}}
\put(-0.7,1.5){\line(-1,1){0.5}}
\put(-0.7,2.0){\line(-1,1){0.5}}
\put(-0.7,2.5){\line(-1,1){0.5}}
\put(-0.7,0.5){\line(0,1){2}}
\put(-0.825,0.625){\line(0,1){2}}
\put(-0.95,0.75){\line(0,1){2}}
\put(-1.075,0.875){\line(0,1){2}}
\put(-1.2,1.0){\line(0,1){2}}
\put(1.3,0.5){\line(-1,1){0.5}}
\put(1.3,1.0){\line(-1,1){0.5}}
\put(1.3,1.5){\line(-1,1){0.5}}
\put(1.3,2.0){\line(-1,1){0.5}}
\put(1.3,2.5){\line(-1,1){0.5}}
\put(1.3,0.5){\line(0,1){2}}
\put(1.175,0.625){\line(0,1){2}}
\put(1.05,0.75){\line(0,1){2}}
\put(0.925,0.875){\line(0,1){2}}
\put(0.8,1.0){\line(0,1){2}}
\put(2.3,0.5){\line(-1,1){0.5}}
\put(2.3,1.0){\line(-1,1){0.5}}
\put(2.3,1.5){\line(-1,1){0.5}}
\put(2.3,2.0){\line(-1,1){0.5}}
\put(2.3,2.5){\line(-1,1){0.5}}
\put(2.3,0.5){\line(0,1){2}}
\put(2.175,0.625){\line(0,1){2}}
\put(2.05,0.75){\line(0,1){2}}
\put(1.925,0.875){\line(0,1){2}}
\put(1.8,1.0){\line(0,1){2}}
\put(-2,3.2){\makebox(0,0){\small $S\delta(z - z_1)$}}
\put(-0.7,0.2){\makebox(0,0){\small $S\delta(z - z_2)$}}
\put(0,1.5){\makebox(0,0){$\ldots$}}
\put(0.8,3.2){\makebox(0,0){\small $S\delta(z - z_{N-1})$}}
\put(2.3,0.2){\makebox(0,0){\small $S\delta(z - z_{N})$}}\end{picture}}
\subfigure[\ Surface roughness]{\includegraphics[scale=0.55]{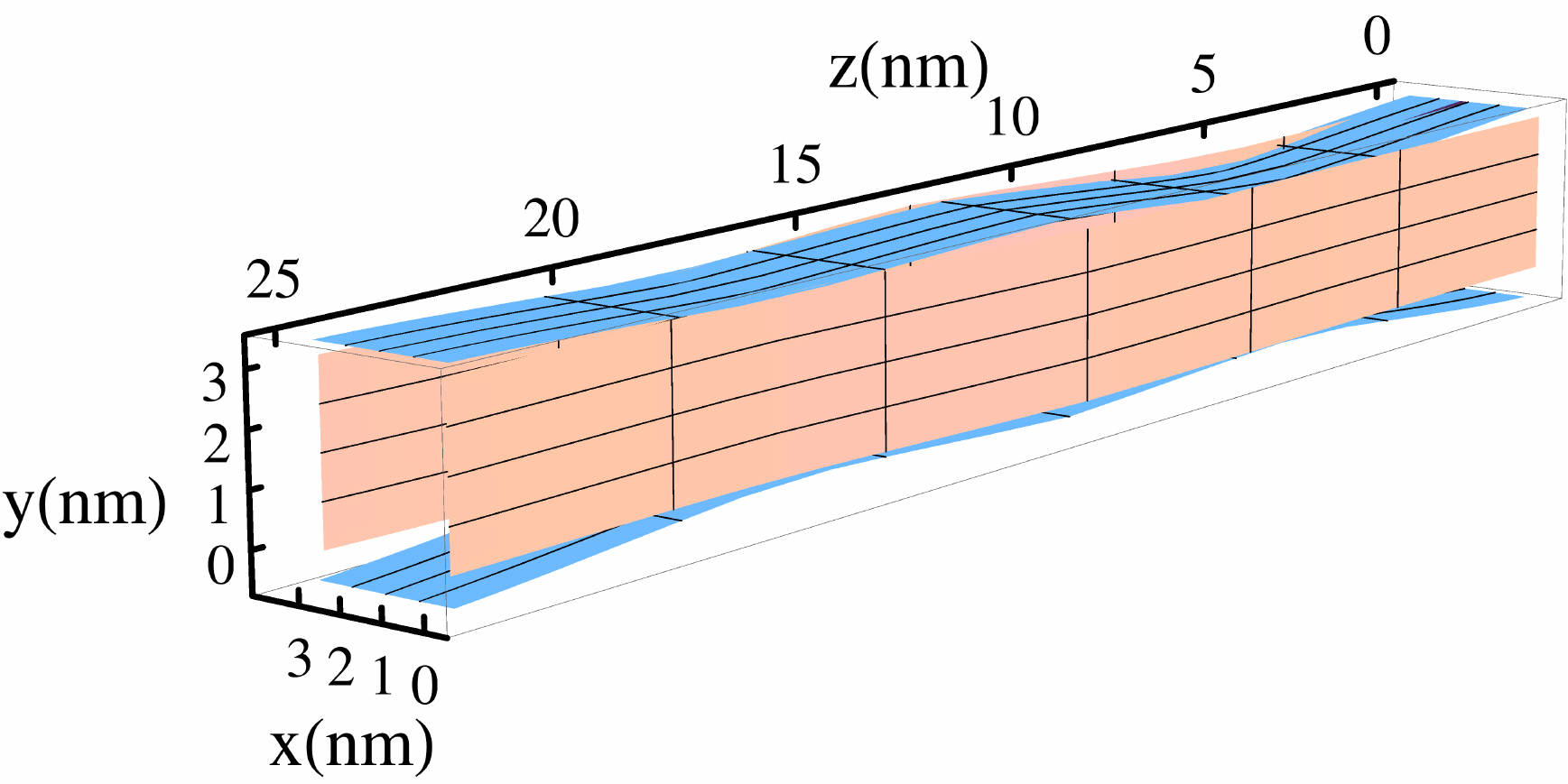}}
\caption{(a) The ideal nanowire without surface roughness and grain-boundaries is modeled as a rectangular box with cross-section sides equal to $D$ and length $L_z$. The electric field $E_z$ and current $J_z$ are along the $z$-direction. Grain-boundaries are also shown, and are always supposed to be infinitely thin and normal to the transport direction. (b) A wire nine copper atoms wide and high (9$a_\textnormal{Cu} \approx 3.3$ nm) is shown here with Gaussian correlated, rough surfaces. The standard deviation is $\Delta = 10\%D$ and the correlation length is $\Lambda = 1.75 D$.}
\label{wireModel}
\end{figure}
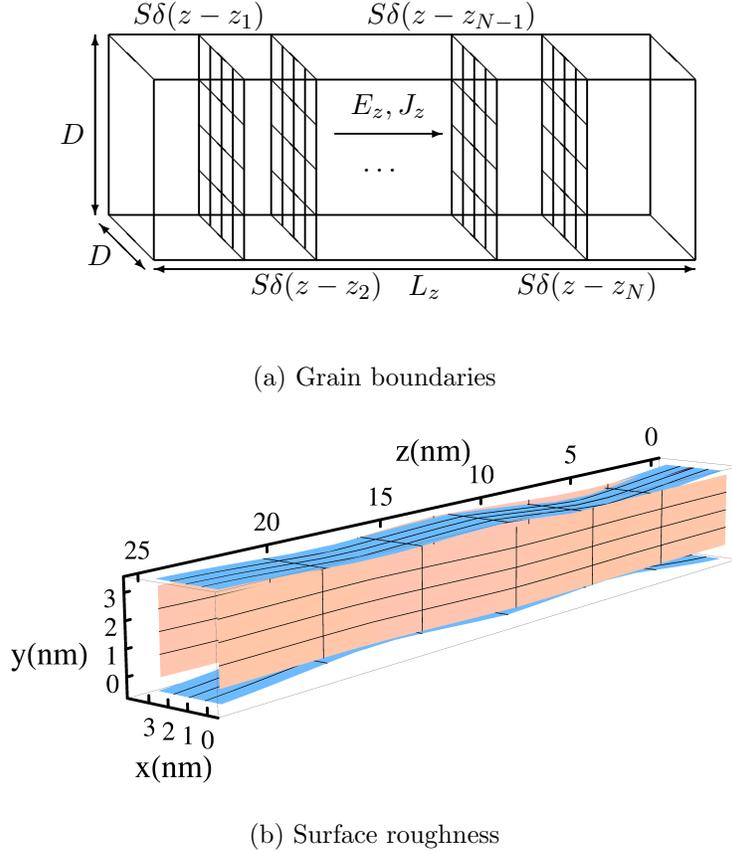

We model the conduction electrons in an ultra-thin metal nanowire as free electrons in an ideal, rectangular box as shown in Fig.~\ref{wireModel} (a). We do not consider cross-section aspect ratios different from one. The box has zero potential inside ($U=0$) and its transverse boundaries are assumed to be hard walls ($U=+\infty$). In the transport direction, denoted by $z$, we impose periodic boundary conditions and the length of the wire is always considered to be long enough, so that the eigenstates can be labeled by a quasi-continuous valued momentum variable $k_z$. Due to the confinement in the transverse directions, denoted by $x$ and $y$, the eigenstates are also labeled by positive integers $n_x$, $n_y$ and have a total energy $E_{n_x n_y}\lef k_z \rig = \hbar^2/\lef 2m^*_e \rig \left[ k_z^2 + \lef \pi n_x/D \rig^2 + \lef \pi n_x/D \rig^2 \right]$. The quadratic expression of the latter reflects the effective mass approximation (EMA) which has been adopted throughout the paper. Although questionable in general for treating nanowires, we have assumed that the EMA provides an acceptable description of the conduction band of a nanowire made of a simple metal like Cu. In particular, we have taken $m^*_e$ to be equal to the free electron mass.

The basis for our model and resistivity calculations is the Boltzmann transport equation (BTE) (see for example Mahan\cite{mahan2000many}), which provides the time evolution of the occupation probability distribution $f_{n_x n_y } \lef k_z \rig$ in phase space, valued between 0 and 1. We do not consider $z$-dependence because a homogeneous electric field $E_z$ is applied. Summing the distribution function over phase space yields the total density of conduction electrons $n_e$ in the wire:
\begin{align*}
n_e &= \frac{2}{2\pi D^2}\sum_{n_x, n_y} \int\limits_{-\infty}^{+\infty} \mkern-5mu \deriv k_z \; f_{n_x n_y}\lef k_z \rig \\
&\stackrel{T=0\textnormal{, eq.}}{=} \frac{2\sqrt{2m_e^*}}{\pi D^2 \hbar} \sum_{n_x, n_y} \sqrt{\textnormal{Max}\left\{0, E_\textnormal{F} - E_{n_x n_y} \right\}},
\end{align*}
with $E_{n_x n_y} \equiv E_{n_x n_y} \lef k_z = 0 \rig$. The second line gives the electron density for the equilibrium distribution at zero temperature, which is used to calculate the Fermi energy $E_\textnormal{F}$ for every diameter of the simulated nanowires. The equilibrium distribution can be used because our perturbative approach to solve for the stationary state distribution function barely changes the resulting Fermi level. The electron density $n_e$ is always fixed to the bulk value of the metal considered. Because of the sub-band quantization, there is a substantial increase of the Fermi energy compared to bulk for few nanometer diameter wires.

The BTE for a stationary state and the resulting expression for the current $J_z$ are given by:
\begin{align}
\label{BTE}
-\frac{e E_z}{\hbar} \frac{\partial f_{n_x n_y}\lef k_z \rig}{\partial k_z} = \left.\frac{\partial f_{n_x n_y}\lef k_z \rig}{\partial t}\right|_\textnormal{collisions}, \\
\label{current}
J_z = -\frac{e}{\pi} \sum_{n_x, n_y} \int\limits_{-\infty}^{+\infty} \mkern-5mu \deriv k_z \; \frac{ \hbar k_z }{m^*_e} f_{n_x n_y} \lef k_z \rig.
\end{align}
Note that a factor of 2 is included in the expression for the current due to the spin degeneracy. The collision term is not specified for the moment, but this will be derived based on quantum-mechanical perturbation theory below. The most important thing that remains to get the current and conductivity with the BTE is to solve for a stationary solution of Eq.~(\ref{BTE}) perturbatively, meaning for small electric field $E_z$ and a close to equilibrium distribution function:
\begin{align}
\label{perturbativeForm}
f_{n_x n_y}\lef k_z \rig \approx f^{(eq.)}_{n_x n_y}\lef k_z \rig + f^{(1)}_{n_x n_y}\lef k_z \rig.
\end{align}
Inserting the typical perturbative expression of the collision term, the RTA:
\begin{align}
\label{RTA}
\left. \frac{\partial f^{(1)}_{n_x n_y} \lef k_z \rig}{\partial t} \right|_\textnormal{collisions} &= - \frac{f^{(1)}_{n_x n_y} \lef k_z \rig}{\tau_{n_x n_y}\lef k_z \rig},
\end{align}
in Eq.~(\ref{BTE}), we get the following solution:
\begin{align}
\notag
f^{(1)}_{n_x n_y}\lef k_z \rig &= -eE_z\tau_{n_x n_y}\lef k_z \rig \frac{\hbar k_z}{m_e} \delta\left[ E_{n_x n_y} \lef k_z \rig - E_\textnormal{F} \right] \\ \label{solutionBTE}
&= -\frac{eE_z}{\hbar} \sum_{\pm} \pm \tau_{n_x n_y}^\pm \delta\lef k_z - k_{z,n_x n_y}^\pm \rig.
\end{align}
Because of the Dirac delta function, the probability distribution is only changed for states at the Fermi-level. It means that the important occupation probabilities are these of the positive and negative momentum states at the Fermi level for each sub-band. There are two solutions for each sub-band with momentum $ k_z = \pm \sqrt{k_\textnormal{F}^2 - \lef \pi n_x / D \rig^2 - \lef \pi n_y / D \rig^2}$ that will be labeled $+(-)$ for positive (negative) $k_z$ states, as in Eq.~(\ref{solutionBTE}). The latter is a zero temperature result, but it provides a good approximation at room temperature because $k_\textnormal{B}T \ll E_\textnormal{F}$ for metals.

We plug Eq.~(\ref{solutionBTE}) into the expression for current, Eq.~(\ref{current}), and retrieve the following formula for the conductivity $\sigma \equiv J_z/(D^2 E_z)$:
\begin{align}
\sigma = \frac{e^2}{\pi m_e^* D^2}\sum_{n_x, n_y, \pm} \tau^{\pm}_{n_x n_y} \left| k_{z,n_x n_y}^{\pm} \right|,
\label{RTAconductivity}
\end{align}
where the relaxation times $\tau_{n_x n_y}^\pm$ are to be extracted from Fermi's golden rule. The latter is invoked to calculate the transition probabilities $P\lef \mid i \rangle \rightarrow \mid f \rangle \rig$ emerging in the full-fledged collision term:

\begin{align}
\label{FGR}
\left.\frac{\partial f^\pm_{n_x n_y}}{\partial t}\right|_\textnormal{collisions} &= \sum_{n_x', n_y'} \frac{L_z}{2\pi} \int\limits_{-\infty}^{+\infty} \mkern-5mu \deriv k_z' \; \left\{ f_{n_x' n_y'}\lef k_z' \rig \lef 1 - f_{n_x n_y}^{\pm} \rig P\lef \mid n_x' n_y', k_z' \rangle \rightarrow \mid n_x n_y \pm \rangle \rig \right. \\ \notag
& \qquad \qquad \qquad \qquad \left. - f_{n_x n_y}^\pm \left[ 1 - f_{n_x' n_y'}\lef k_z' \rig \right] P\lef \mid n_x n_y \pm \rangle \rightarrow \mid n_x' n_y', k_z' \rangle \rig \right\}, \\ \label{FGReq}
P\lef \mid i \rangle \rightarrow \mid f \rangle \rig &= \frac{2\pi}{\hbar}\left| \langle i \mid V \mid f \rangle \right|^2 \delta\lef E_i - E_f \rig \equiv \frac{2\pi}{\hbar} M_i^f \delta\lef E_i - E_f \rig,
\end{align}

expressing the change of the occupation probability due to collisions as the sum over all the probabilities to scatter in from every other state and the sum over all the probabilities to scatter out from the state we are considering.\footnote{There is no sum over spin states, because a spin flip cannot be induced by surface roughness or grain-boundaries. In principle this analysis could be regarded as the solution for one of the two spin states. The solution for the other spin state is completely analogous, hence a factor 2 in the conductivity expression using the probability distribution function neglecting the spin.}

The perturbation Hamiltonian term $V$ describes the difference between the realistic nanowire, with surface roughness and grain-boundaries, and the ideal rectangular box Hamiltonian. The calculation of these matrix elements $M_{n_x n_y, k_z}^{n_x' n_y', k_z'}$ for surface roughness scattering is discussed in section \ref{sectionSRS} whereas section \ref{sectionGBS} covers grain-boundary scattering.

The transition probabilities $P$ in Eq.~(\ref{FGR}), defining a typical time scale before a state is scattered into another state, obey conservation of energy. Hence, only states at the Fermi level participate in the collision term, which is consistent with the zero temperature assumption.

We see from Eq.~(\ref{FGReq}) that the transition probabilities are symmetric under interchange of initial and final state. One of the consequences is that the exclusion-blocked transitions for incoming and outgoing scattering are completely identical and have no effect on the collision term. Using Eq.~(\ref{solutionBTE}), (\ref{FGR}) and (\ref{FGReq}), we obtain:
\begin{align}
1 & = \frac{m_e L_z}{\hbar^3}\sum_{\stackrel{\left\{ n_x', n_y', \pm' \right\}}{\neq \left\{ n_x, n_y, \pm \right\}}} \lef - \frac{k_{z,n_x' n_y'}^{\pm'}}{k_{z,n_x n_y}^\pm} \tau_{n_x' n_y'}^{\pm'} + \tau_{n_x n_y}^{\pm} \rig \frac{M_{n_x n_y \pm}^{n_x' n_y' \pm'}}{\left| k_{z,n_x' n_y'}^{\pm'} \right|}.
\label{coupledEquations}
\end{align}
Note that in the above system of equations all relaxation times are coupled. The most crude approximation to retrieve $\tau_{n_x n_y}^\pm$ and get the conductivity via Eq.~(\ref{RTAconductivity}) would neglect incoming scattering, hence putting the first term on the right-hand side equal to zero, yielding:
\begin{align}
\label{noInScattering}
& \left.\frac{1}{\tau_{n_x n_y}^\pm}\right|_\textnormal{No in-scattering} \equiv \frac{m_e L_z}{\hbar^3} \sum_{n_x', n_y', \pm'} \frac{M_{n_x n_y, k_z}^{n_x' n_y', k_z'}}{\left| k_{z, n_x' n_y'}^{\pm'} \right|}.
\end{align}
A better approximation to decouple the equations, is retrieved assuming that the relaxation times are the same for every sub-band state at the Fermi level. In this case one gets the following closed form expression for the relaxation time:
\begin{align}
\label{equalRT}
& \left.\frac{1}{\tau_{n_x n_y}^\pm}\right|_\textnormal{Equal RT} \equiv \frac{m_e L_z}{\hbar^3} \sum_{n_x', n_y', \pm'} \frac{M_{n_x n_y, k_z}^{n_x' n_y', k_z'}}{\left| k_{z, n_x' n_y'}^{\pm'} \right|} \lef 1 - \frac{k_{z, n_x' n_y'}^{\pm'}}{k_{z, n_x n_y}^\pm} \rig.
\end{align}
Note that this approximation is also underlying the well-known bulk form of the relaxation time,\cite{ashcroft1976solid} $\tau\lef \mathbf{k} \rig^{-1}_\textnormal{Bulk} = \sum_{\mathbf{k}'} P\lef \mid \mathbf{k} \rangle \rightarrow \mid \mathbf{k'} \rangle \rig \lef 1 - \hat{\mathbf{k}} \cdot \hat{\mathbf{k}}' \rig$. This approximation has been used previously to incorporate incoming scattering,\cite{jin2007modeling} but its validity is not guaranteed and one can easily see that the effective relaxation time could become negative in Eq.~(\ref{equalRT}).

The correct way to solve the BTE perturbatively, is by solving the complete set of coupled equations in Eq.~(\ref{FGR}) for all the different relaxation times. Because it is a system of linear equations, finding the solution just boils down to the inversion of a large square matrix, i.e.:
\begin{align}
&\lef \begin{matrix}
\tau_{1 1}^{+} \\
\tau_{1 2}^{+} \\
\vdots \\
\tau_{n_x^* n_y^*}^{+} \\
\end{matrix} \rig \label{exactSolutionMatrix} = \\ \notag
& \quad \lef \begin{matrix}
\left. \frac{1}{\tau_{1 1}^+} \right|_{\textnormal{\tiny No in-sc.}} + \frac{m_e L_z M_{1 1 +}^{1 1 -}}{\hbar^3 k_{z,1 1}^+ } & - \frac{m_e L_z \sum_\pm \pm M_{1 1 +}^{1 2 \pm}}{\hbar^3 k_{z, 1 1}^+} & \cdots & - \frac{m_e L_z \sum_\pm \pm M_{1 1 +}^{n_x^* n_y^* \pm}}{\hbar^3 k_{z, 1 1}^+} \\
- \frac{m_e L_z \sum_\pm \pm M_{1 2 +}^{1 1 \pm}}{\hbar^3 k_{z, 1 2}^+} & \left. \frac{1}{\tau_{1 2}^+} \right|_{\textnormal{\tiny No in-sc.}} + \frac{m_e L_z M_{1 2 +}^{1 2 -}}{\hbar^3 k_{z,1 2}^+ } & \cdots & - \frac{m_e L_z \sum_\pm \pm M_{1 2 +}^{n_x^* n_y^* \pm}}{\hbar^3 k_{z, 1 2}^+} \\
\vdots & \vdots & \ddots & \vdots \\
- \frac{m_e L_z \sum_\pm \pm M_{n_x^* n_y^* +}^{1 1 \pm}}{\hbar^3 k_{z, n_x^* n_y^*}^+} & - \frac{m_e L_z \sum_\pm \pm M_{n_x^* n_y^* +}^{1 2 \pm}}{\hbar^3 k_{z, n_x^* n_y^*}^+} & \cdots & \left. \frac{1}{\tau_{n_x^* n_y^*}^+} \right|_{\textnormal{\tiny No in-sc.}} + \frac{m_e L_z M_{n_x^* n_y^* +}^{n_x^* n_y^* -}}{\hbar^3 k_{z,n_x^* n_y^*}^+ }
\end{matrix} \rig^{-1} \lef \begin{matrix}
1 \\
1 \\
\vdots \\
1
\end{matrix} \rig.
\end{align}

Note that only positive momentum states need to be considered, as the opposite momenta yield the same relaxation times. We have introduced the labels $n_x^*,n_y^*$ to denote the highest integer values corresponding with the highest sub-band having states below the Fermi-energy. Denoting the matrix inverse in Eq.~(\ref{exactSolutionMatrix})  by $\mathcal{T}_{n_x n_y, n_x' n_y'}$, we can obtain the correct state-dependent relaxation times $\tau_{n_x n_y}^\pm$ without having approximated the solution of the perturbed BTE:
\begin{align}
\label{exactRT}
\tau_{n_x n_y}^\pm &= \sum_{n_x' n_y'} \mathcal{T}_{n_x n_y, n_x' n_y'}.
\end{align}
These relaxation times can be interpreted as the state-dependent lifetimes, such that a sub-band quantized generalization of the Drude conductivity appears in Eq.~(\ref{RTAconductivity}). Results of the correct relaxation times and resulting conductivities, compared to the approximated forms [\textit{No in-scattering}, Eq.~(\ref{noInScattering}), and \textit{Equal RT}, Eq.~(\ref{equalRT})] are shown in section \ref{results}.

\section{Matrix elements}

\subsection{Surface roughness}
\label{sectionSRS}
To model the surface roughness, we first introduce four functions providing the fluctuations around the flat, ideal boundaries wire: $S_{x=0}(y,z)$, $S_{x=D}(y,z)$, $S_{y=0}(x,z)$ and $S_{y=D}(x,z)$. Typically one supposes Gaussian or exponential autocorrelation functions to model the boundary roughness:
\begin{align*}
\left< S_{x=0}(y,z) S_{x=0}(y',z') \right> &= \Delta^2 e^{-\frac{(y - y')^2 + (z-z')^2}{\Lambda^2/2} } \textnormal{(Gaussian)}, \\
\left< S_{x=0}(y,z) S_{x=0}(y',z') \right> &= \Delta^2 e^{-\frac{\sqrt{(y - y')^2 + (z-z')^2}}{\Lambda/\sqrt{2}} } \textnormal{(exp.)},
\end{align*}
with standard deviation $\Delta$ and correlation length $\Lambda$. We will calculate the scattering matrix elements for Gaussian correlated rough surfaces.

The matrix elements $\langle n_x n_y \pm \mid V_\textnormal{SR} \mid n_x' n_y' \pm' \rangle$ are given by:
\begin{align}
\label{matrixElement}
\langle n_x n_y \pm \mid V_\textnormal{SR} \mid n_x' n_y' \pm' \rangle &= \frac{1}{L_z}\int\limits_0^{D} \deriv x \int\limits_0^{D} \deriv y \mkern-8mu \int\limits_{-L_z/2}^{+L_z/2} \mkern-10mu \deriv z \; \psi^*_{n_x}\lef x \rig \psi^*_{n_y}\lef y \rig e^{-ik^\pm_{z,n_x n_y} z} \\ \notag
& \quad \times \left[ H(x,y,z) - H_0(x,y,z) \right] \psi_{n_x'}\lef x \rig \psi_{n_y'}\lef y \rig e^{ik^{\pm'}_{z,n_x' n_y'} z},
\end{align}
with the approximated Hamiltonian $H_0$ given by:
\begin{align*}
H_0 \lef \mathbf{r} \rig &=
\left\{
\begin{matrix}
-\frac{\hbar^2}{2m_e}\nabla^2, \qquad \quad \mathbf{r} \in \left[ 0, D \right] \times \left[ 0, D \right] \times \left[ -L_z/2, +L_z/2 \right] \\
-\frac{\hbar^2}{2m_e}\nabla^2+U, \quad \mathbf{r} \notin \left[ 0, D \right] \times \left[ 0, D \right] \times \left[ -L_z/2, +L_z/2 \right]
\end{matrix} \right. ,
\end{align*}
and $\psi_{n_x}(x) = (2/D)^{1/2} \sin\lef n_x \pi x / D \rig$. The difference between the correct and approximate Hamiltonian is only non-zero near the boundaries that are shifted due to roughness. This difference diverges if the potential well height $U$ is infinitely high, rendering the integral in Eq.~(\ref{matrixElement}) divergent. We can however expand Eq.~(\ref{matrixElement}) up to first order of the roughness function $S_{x=0}(y,z)$, which yields a finite result, first introduced by Prange and Nee.\cite{prange1968quantum} The resulting matrix element for the $x=0$ surface reads:
\begin{align}
\label{matrixElementInfiniteWell}
& \langle n_x n_y \pm \mid V_{SR, x=0} \mid n_x' n_y' \pm' \rangle \\ \notag
& \quad \approx 2\frac{\sqrt{E_{n_x} E_{n_x'}}}{D L_z} \int\limits_0^{D} \deriv y \mkern-8mu \int\limits_{-L_z/2}^{+L_z/2} \mkern-10mu \deriv z \; \psi^*_{n_y}\lef y \rig e^{-ik^\pm_{z,n_x n_y} z} S_{x=0}(y,z) \psi_{n_y'}\lef y \rig e^{ik^{\pm'}_{z,n_x' n_y'} z}.
\end{align}
Averaging the squared absolute value appearing in Fermi's golden rule with the surface roughness autocorrelation functions, we arrive at:
\begin{align}
\label{SRSMatrixElement}
&\left< \left| \langle n_x n_y \pm \mid V_{SR, x=0} \mid n_x' n_y' \pm' \rangle \right|^2 \right>_{S_{x=0}} \\ \notag
& \quad \approx 4 \frac{E_{n_x} E_{n_x'}}{D^2} \frac{\Delta^2}{L_z^2} \int\limits_0^{D} \deriv y \mkern-8mu\int\limits_{-L_z/2}^{+L_z/2} \mkern-10mu \deriv z \int\limits_0^{D} \deriv y' \mkern-8mu \int\limits_{-L_z/2}^{+L_z/2} \mkern-10mu \deriv z' \; \psi^*_{n_y}\lef y \rig \psi_{n_y}\lef y' \rig \psi_{n_y'}\lef y \rig \psi_{n_y'}^*\lef y' \rig \\ \notag
& \quad \quad \times \exp\left[ -ik^\pm_{z,n_x n_y} z + ik^\pm_{z,n_x n_y} z' + ik^{\pm'}_{z,n_x' n_y'} z - ik^{\pm'}_{z,n_x' n_y'} z' - \frac{2}{\Lambda^2}(y - y')^2 - \frac{2}{\Lambda^2}(z-z')^2 \right],
\end{align}
for the Gaussian model. The integral can be solved analytically, as its integrand is a product of exponentials of terms that are only linear and quadratic in $y$, $y'$, $z$ and $z'$. In practice, we can reduce the analytical expression by dropping terms that disappear when the correlation length is much smaller than $L_z$. We can however not make a similar simplification for $D$, as typical values of $\Lambda$ in nanowires fabricated today have correlation lengths comparable to the diameter. Note that the total matrix element value is just the sum of the contribution of all the different surfaces, because we suppose that the surfaces are not cross-correlated.

\subsection{Grain-boundaries}
\label{sectionGBS}
The interaction potential representing $N$ grain-boundaries in the nanowire is given by a series of delta functions,
\begin{align}
\label{GBPotential}
V_\textnormal{GB}(x,y,z) = \sum_{\alpha = 1}^{N}U_\textnormal{GB} L_\textnormal{GB} \delta\lef z - z_\alpha \rig,
\end{align}
where the grain-boundary planes $\lef z = z_\alpha \rig$ are perpendicular to the transport direction and represent the misaligned planes of the crystal grains. The strength of the grain-boundary potential is represented by $U_\textnormal{GB} L_\textnormal{GB}$, having the dimensions of energy times distance, the two factors respectively representing the energy barrier height of the grain-boundary and its width. The positions of the grain-boundaries, $z_\alpha$, are distributed according to a Gaussian distribution $g\lef z_1, \ldots, z_N \rig$,
\begin{align*}
g\lef z_1, \ldots, z_N \rig = \frac{\exp\left[-\sum\limits_{\alpha = 1}^{N-1} \lef z_{\alpha+1} - z_\alpha - \frac{L_z}{N} \rig^2/2\sigma_\textnormal{\tiny GB}^2\right]}{L_z \lef 2\pi \sigma_\textnormal{\tiny GB}^2 \rig^{(N-1)/2}}.
\end{align*}
The grain-boundaries are on average uniformly distributed along the wire, with standard deviation $\sigma_\textnormal{\tiny GB}$. The squared matrix element, averaged over the grain-boundary positions, has a non-zero result only for Umklapp scattering because of the definition of the potential in Eq.~(\ref{GBPotential}):

\begin{align*}
& \left< \left| \langle n_x n_y \pm \mid V_\textnormal{GB} \mid n_x n_y \mp \rangle \right|^2 \right>_{z_\alpha} \\
& \quad = \frac{U_\textnormal{GB}^2 L_\textnormal{GB}^2}{L_z^2} \sum_{\alpha, \alpha'} \left< e^{- i 2 \lef k_{z,n_x n_y}^\pm \rig \lef z_\alpha - z_{\alpha'} \rig} \right>_{z_{1, \ldots, N}} \\
& \quad \approx \frac{U_\textnormal{GB}^2 L_\textnormal{GB}^2}{L_z^2} \sum_{\alpha, \alpha'} \int\limits_{-L_z/2}^{+L_z/2} \mkern-10mu \deriv z_1 \mkern-5mu \int\limits_{-\infty}^{+\infty} \mkern-5mu \deriv z_2 \cdots \int\limits_{-\infty}^{+\infty} \mkern-5mu \deriv z_N \; g\lef z_1, \ldots, z_N \rig e^{- i 2 k_{z,n_x n_y}^\pm \lef z_\alpha - z_{\alpha'} \rig} \\
& \quad \approx \frac{N U_\textnormal{GB}^2 L_\textnormal{GB}^2}{L_z^2} \frac{\sinh\left[ 2 \lef k_{z, n_x n_y}^\pm \sigma_\textnormal{\tiny GB} \rig^2 \right]}{\cosh\left[ 2 \lef k_{z, n_x n_y}^\pm \sigma_\textnormal{\tiny GB} \rig^2 \right] - \cos \left[ 2k_{z, n_x n_y}^\pm L_z/N \right]}.
\end{align*}
In analogy with Mayadas,\cite{mayadas1970electrical} we have made two approximations that affect the result very little as long as $N$ is quite large and $\sigma_\textnormal{\tiny GB} \ll L_z$. In the third line we have extended the integration limits of $z_\alpha$ with $\alpha$ between 2 and $N$ from $-\infty$ to $+\infty$ and in the last line we have neglected the $\mathcal{O}\lef 1/N \rig$ part of the solution. The total matrix element that should be inserted into Eq.~(\ref{FGReq}) is just the sum of the boundary surface contributions and the grain-boundary contributions, because the correlations between boundary surfaces and grain-boundaries appearing in the cross-terms, e.g. $\left< \langle i \mid V_\textnormal{SR} \mid f \rangle \langle f \mid V_\textnormal{GB} \mid i \rangle \right>_{S,z_\alpha}$, are considered to be zero.

\section{Results \& Discussion}
\label{results}

\subsection{Relaxation times}
\label{relaxationTimes}

\begin{figure}[htb]
\centering
\subfigure[\ Surface roughness]{\includegraphics[scale=0.4]{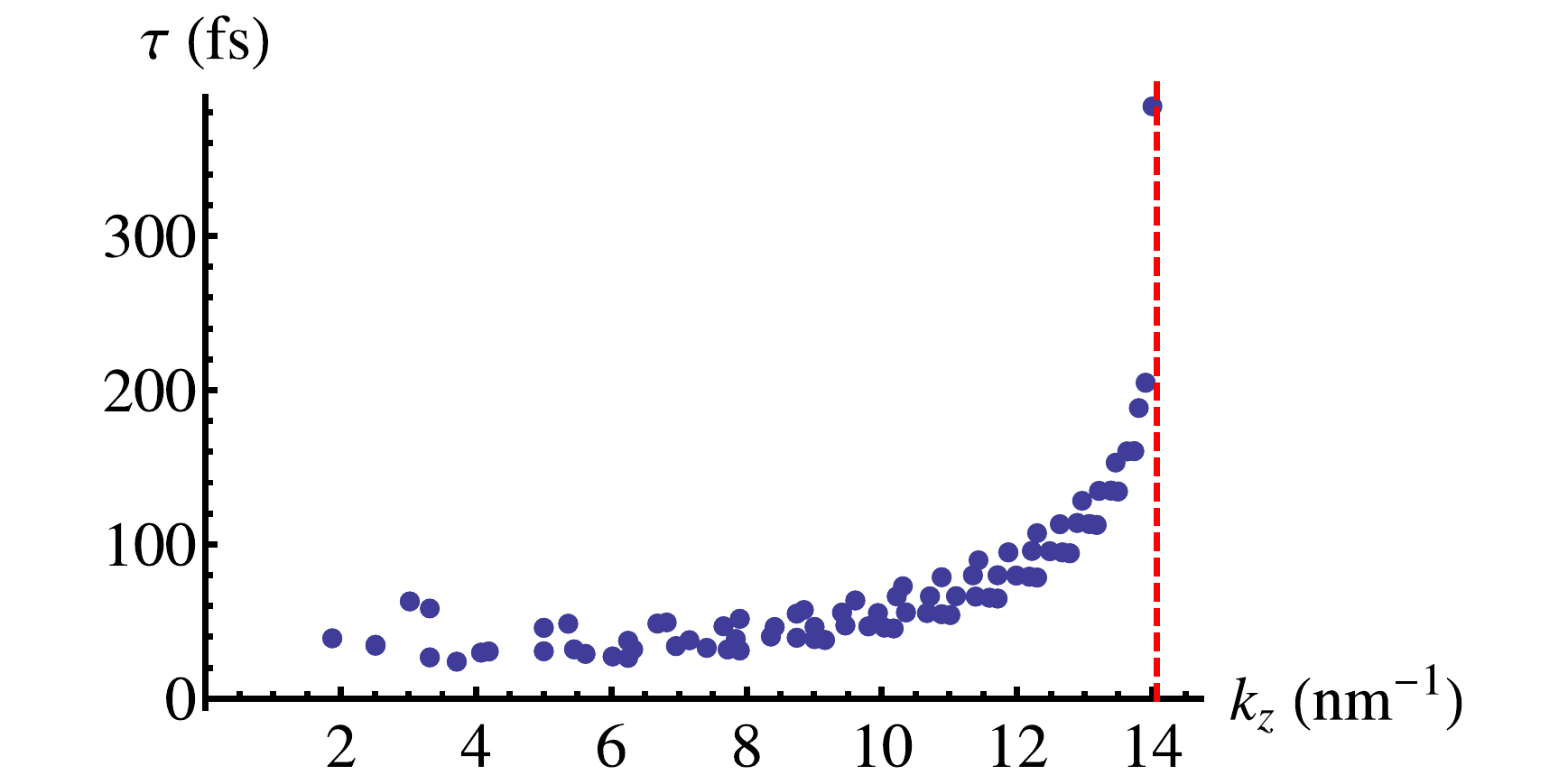}}
\subfigure[\ Grain boundaries]{\includegraphics[scale=0.4]{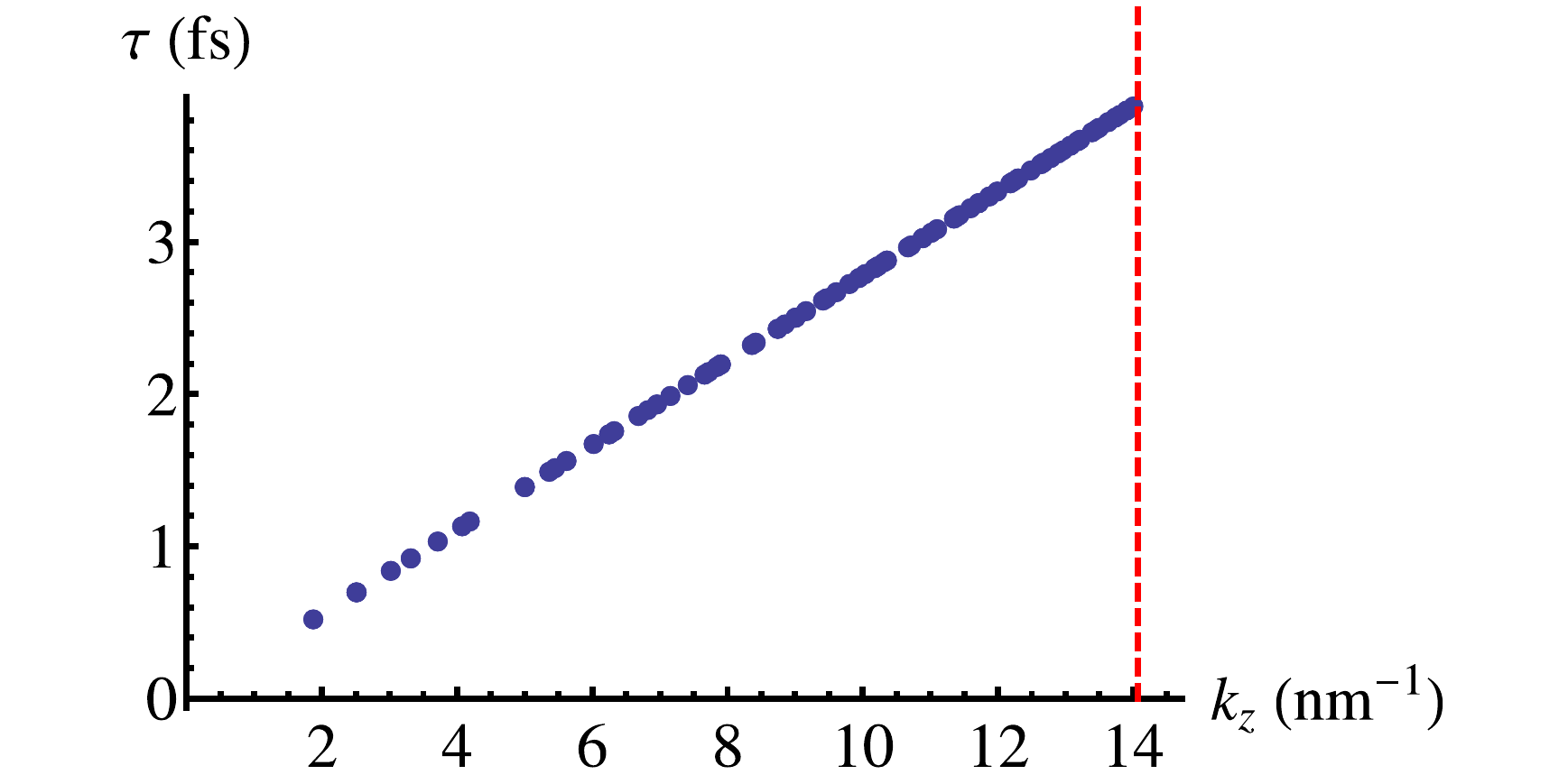}}
\caption{\label{RTPlot} The relaxation times for all Fermi level states of a wire with cross-section of 9 by 9 $a_\textnormal{Cu}$ ($D \approx 3.3$ nm) are shown as a function of their transport momentum $k_z$, limited to positive momenta. Constant Fermi wave vector $k_\textnormal{F}$ is shown as a red, dashed line. The simulated wire has (a) surface roughness with $\Delta = 0.1 D$ and $\Lambda = 10$ nm or (b) grain-boundaries with $U_\textnormal{GB} = 1.5$eV, $L_\textnormal{GB} = a_\textnormal{Cu}$, $L_z/N = D$ and $\sigma_\textnormal{\tiny GB} = L_z/2N$.}
\end{figure}

\begin{figure}[htb]
\centering
\subfigure[\ Surface roughness]{\includegraphics[scale=0.4]{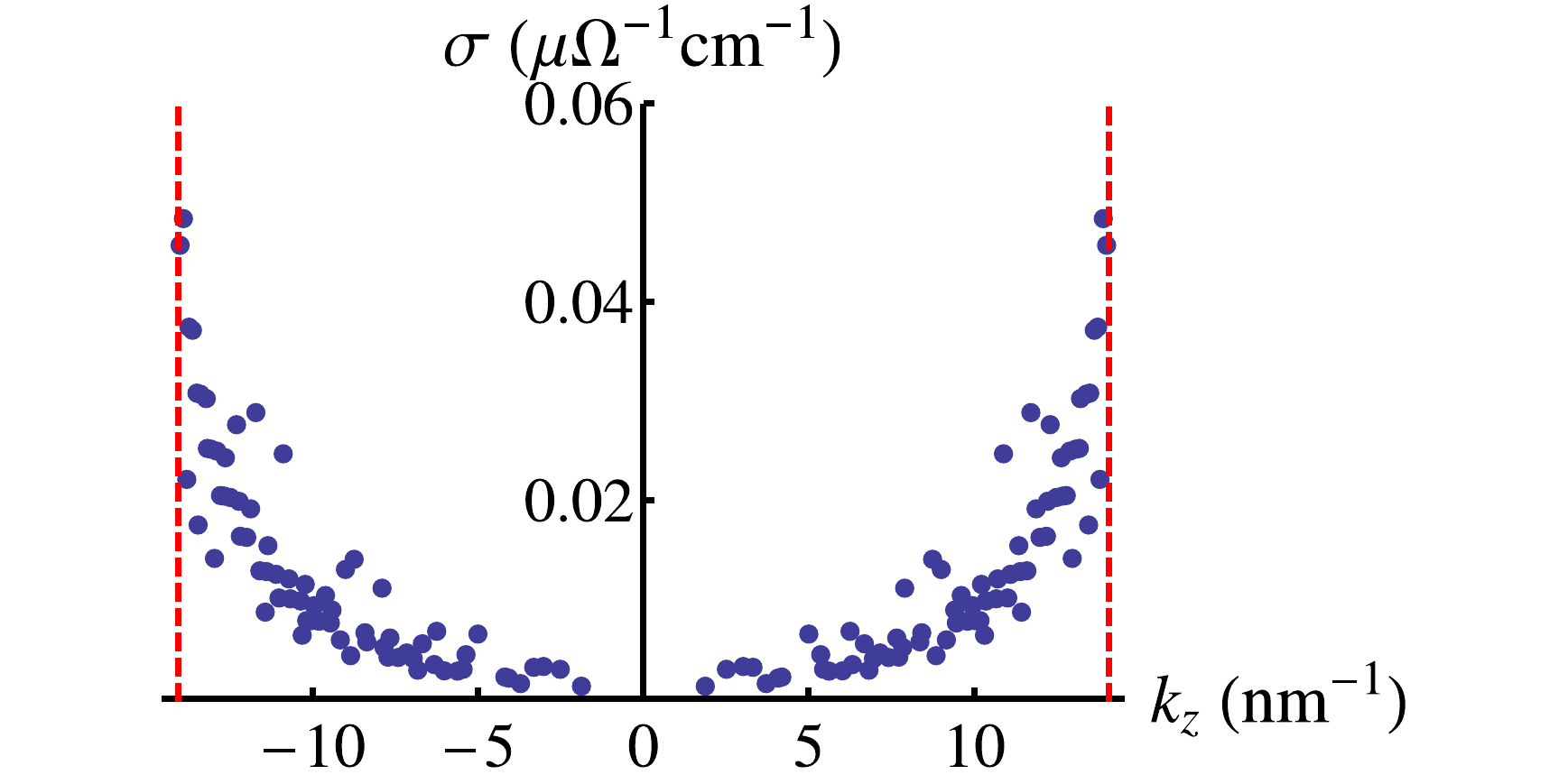}}
\subfigure[\ Grain boundaries]{\includegraphics[scale=0.4]{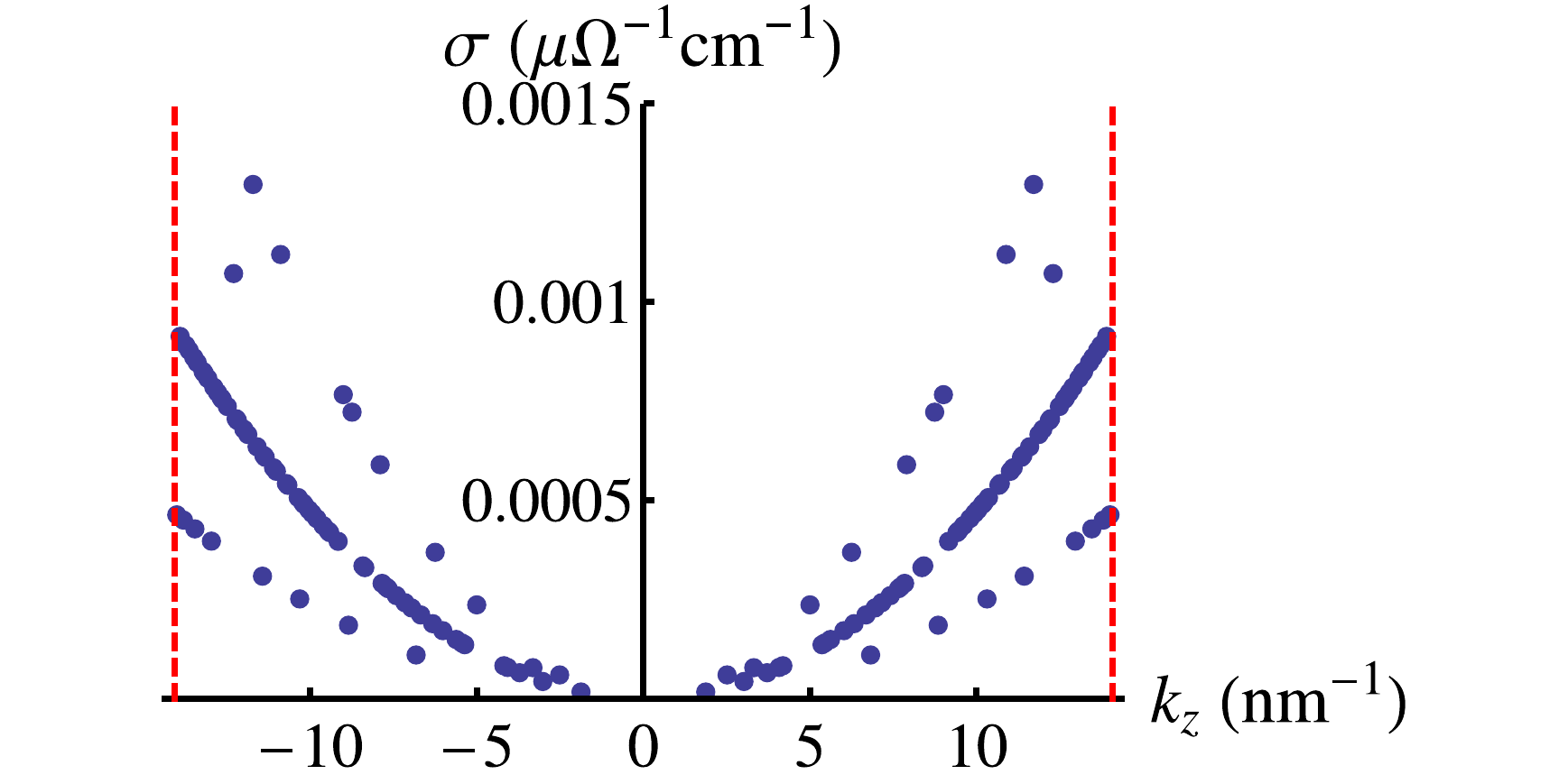}}
\caption{\label{ConductivityPlot} The conductivity contributions of all the Fermi level states as a function of $k_z$ are shown for a copper wire with cross-section of 9 by 9 $a_\textnormal{Cu}$ ($D \approx 3.3$ nm). Constant Fermi wave vector $k_\textnormal{F}$ and $-k_\textnormal{F}$ are shown as red, dashed lines. In case of (a) surface roughness, $\Delta = 0.1 D$ and $\Lambda = 10$ nm are considered, in case of (b) grain-boundaries, $U_\textnormal{GB} = 1.5$eV, $L_\textnormal{GB} = a_\textnormal{Cu}$, $L_z/N = D$ and $\sigma_\textnormal{\tiny GB} = D/2$ are considered.}
\end{figure}

The relaxation times due to surface roughness and grain-boundaries for all Fermi level states, calculated using Eq.~(\ref{exactRT}), are plotted for a copper nanowire with 3.3 nm sides in Fig.~\ref{RTPlot} as a function of $k_z$. Only the relaxation times for the positive momentum states are shown, but the negative momentum states have the same relaxation time as their opposite $k_z$ state. Clearly, the relaxation time is not constant in general, indicating that the equal RT approximation is not a good approximation. Moreover, the relaxation time is largest for the states with the highest transport momentum, such that the conductivity contribution of these states is enhanced both by their high momentum and lifetime.

The relaxation times of surface roughness are strongly peaked for high $k_z$ states. The states with high $n_x$, $n_y$ have more energy in the transverse direction, thus a larger scattering matrix element as can be seen in Eq.~(\ref{SRSMatrixElement}) and a dramatic decrease of their lifetime. There is some spread in the relaxation times because the states with sub-band indices close to $n_x = n_y$ are slightly more stable than the states close to $n_{x/y} = 1$. The relaxation times are in general quite large, of the order of 100 fs. In contrast, the grain-boundary relaxation times are only of the order of 1-10 fs. The high $k_z$ states are also the most stable, but the difference is much less, and can be attributed solely to the difference in the density of states. For the considered values of $\sigma_\textnormal{\tiny GB}$, the matrix elements themselves are almost independent of the states. The correct relaxation times for a nanowire with grain-boundaries and surface roughness are retrieved by adding the matrix element contributions of surface roughness and grain-boundaries together. Because surface roughness is much less important, we study their contribution separately to clearly see the different scaling behavior.

The conductivity contribution to each state as a function of $k_z$ is shown in Fig.~\ref{ConductivityPlot} for the same wire, according to Eq.~(\ref{RTAconductivity}). For grain-boundaries the conductivity is sharply peaked around the Fermi wave vector $k_\textnormal{F}$, much less so for surface roughness. The difference in conductivity for grain-boundary scattering and surface roughness scattering is large, as was expected from the large difference in relaxation times in Fig.~\ref{RTPlot}. The $k_z$ degeneracy of the different states can be clearly recognized in the different lines of the grain-boundary plot.

\subsection{Resistivity scaling}

\begin{figure}[h]
\centering
\subfigure[\ Surface roughness]{\includegraphics[scale=0.5]{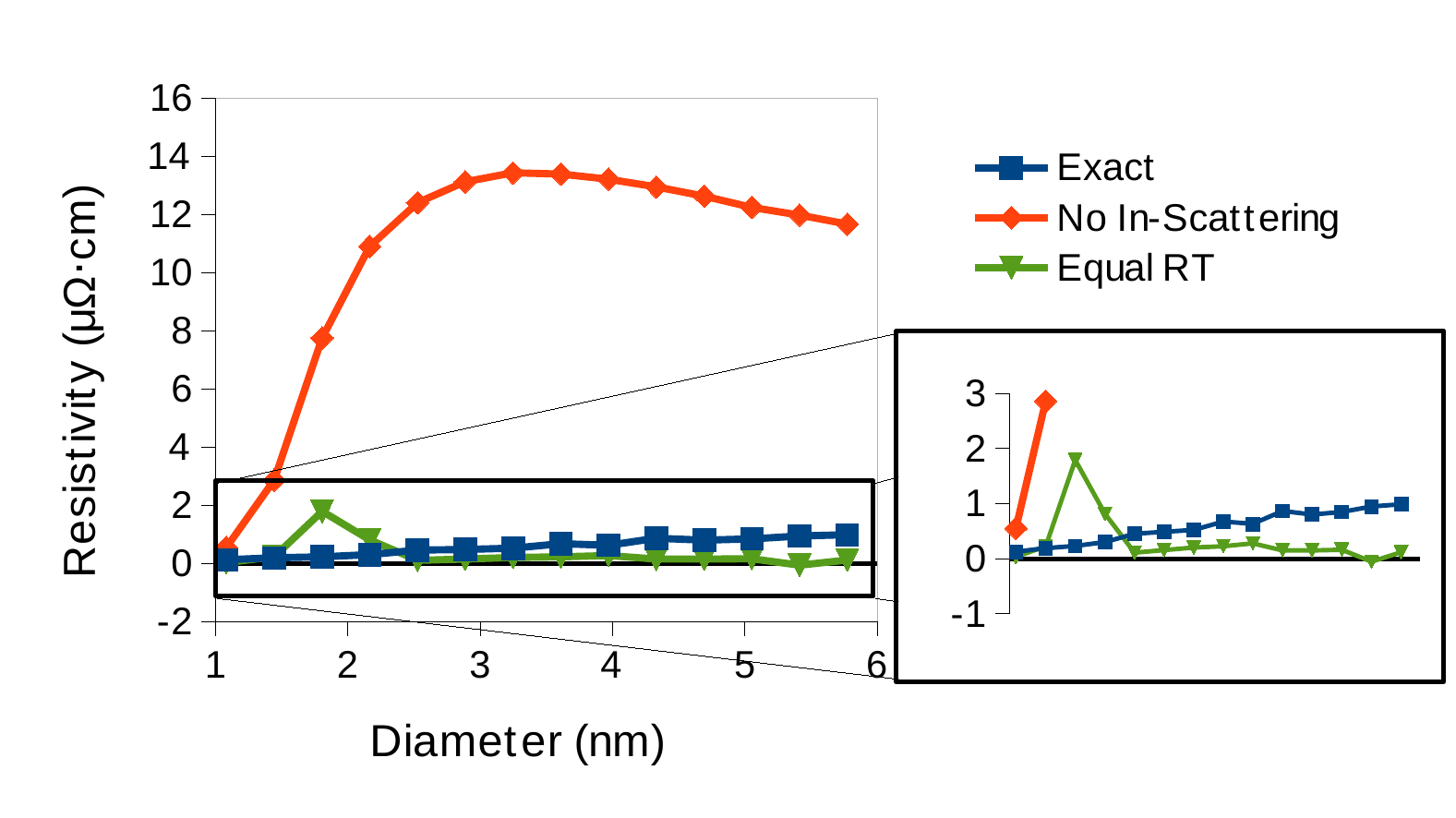}}
\subfigure[\ Grain boundaries]{\includegraphics[scale=0.5]{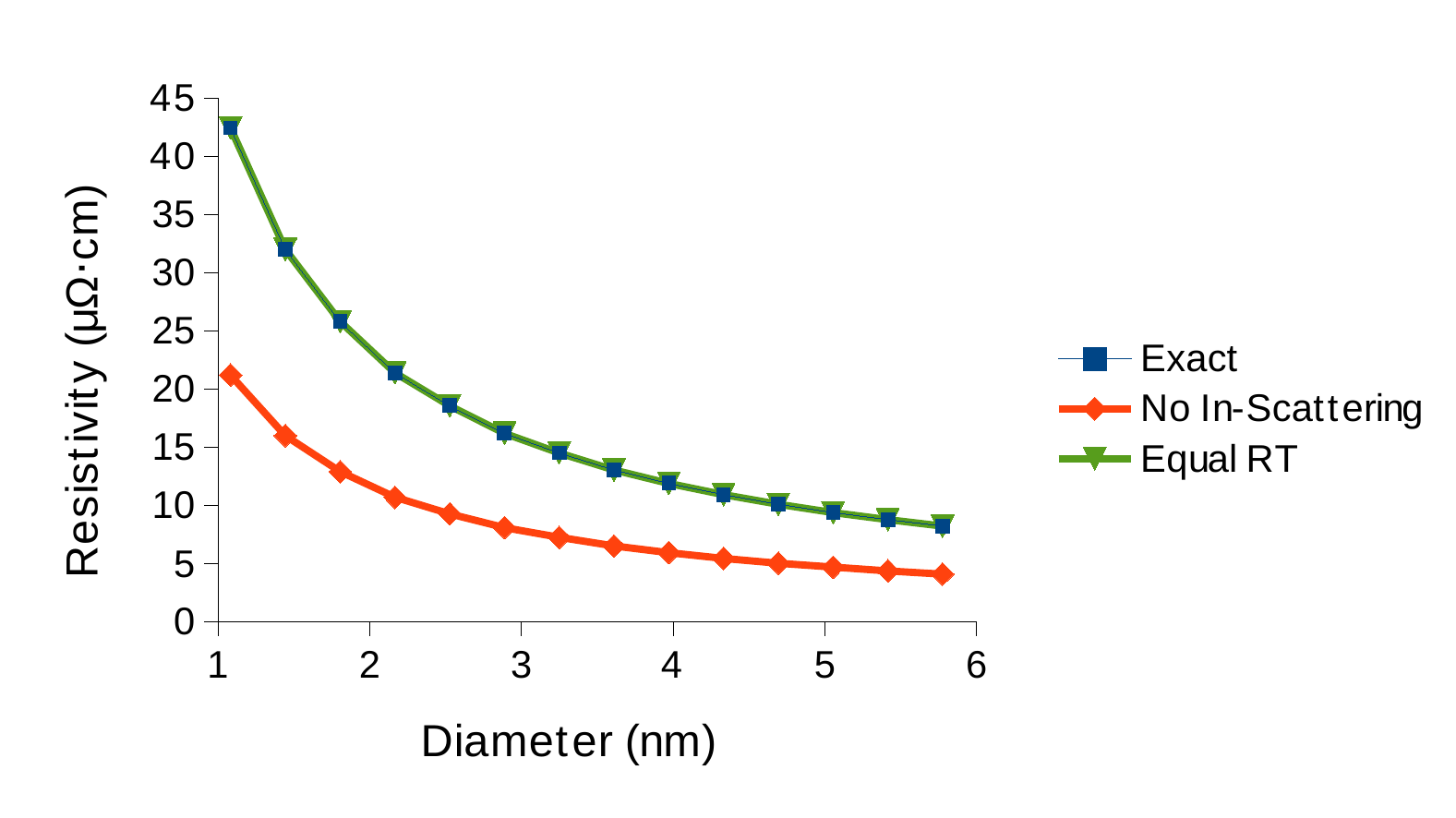}}
\caption{\label{resistivityScaling} The resistivity is shown for copper nanowires of different diameters. The exact result (Exact) is compared to the approximative methods explained in section \ref{RTA} (No In-Scattering and Equal RT). (a) Scaling roughness size: $\Delta = 0.1D$, constant correlation length: $\Lambda \approx 10$ nm (b) Scaling grain-boundary density: $L_z/N = D$, scaling standard deviation: $\sigma_\textnormal{\tiny GB} = D/2$, $U_\textnormal{GB} = 1.5$eV, $L_\textnormal{GB} = a_\textnormal{Cu}$. Note that the \textit{Equal RT} and \textit{Exact} results coincide.}
\end{figure}

\begin{figure}[h]
\centering
\subfigure[\ Surface roughness]{\includegraphics[scale=0.5]{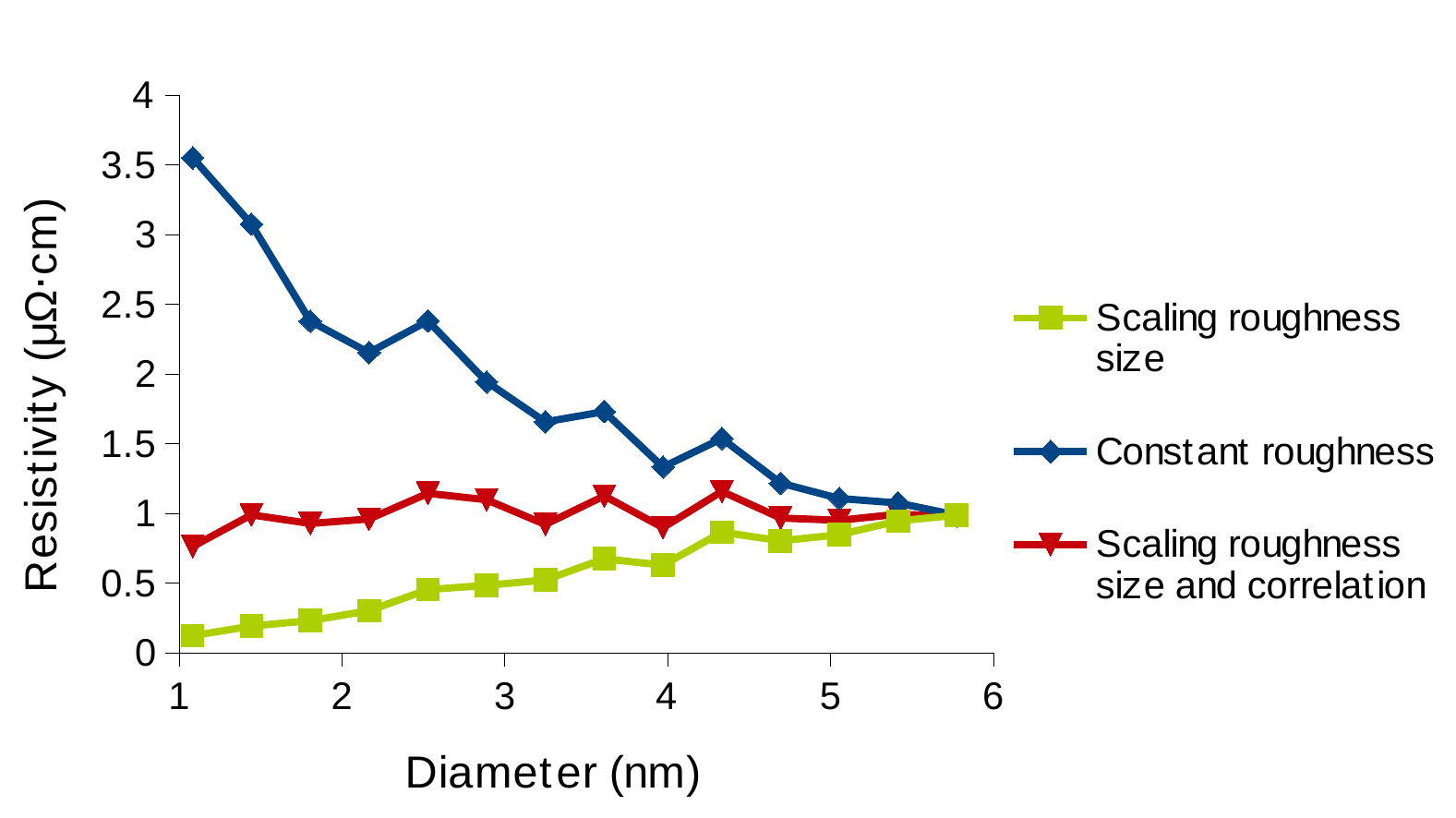}}
\subfigure[\ Grain boundaries]{\includegraphics[scale=0.5]{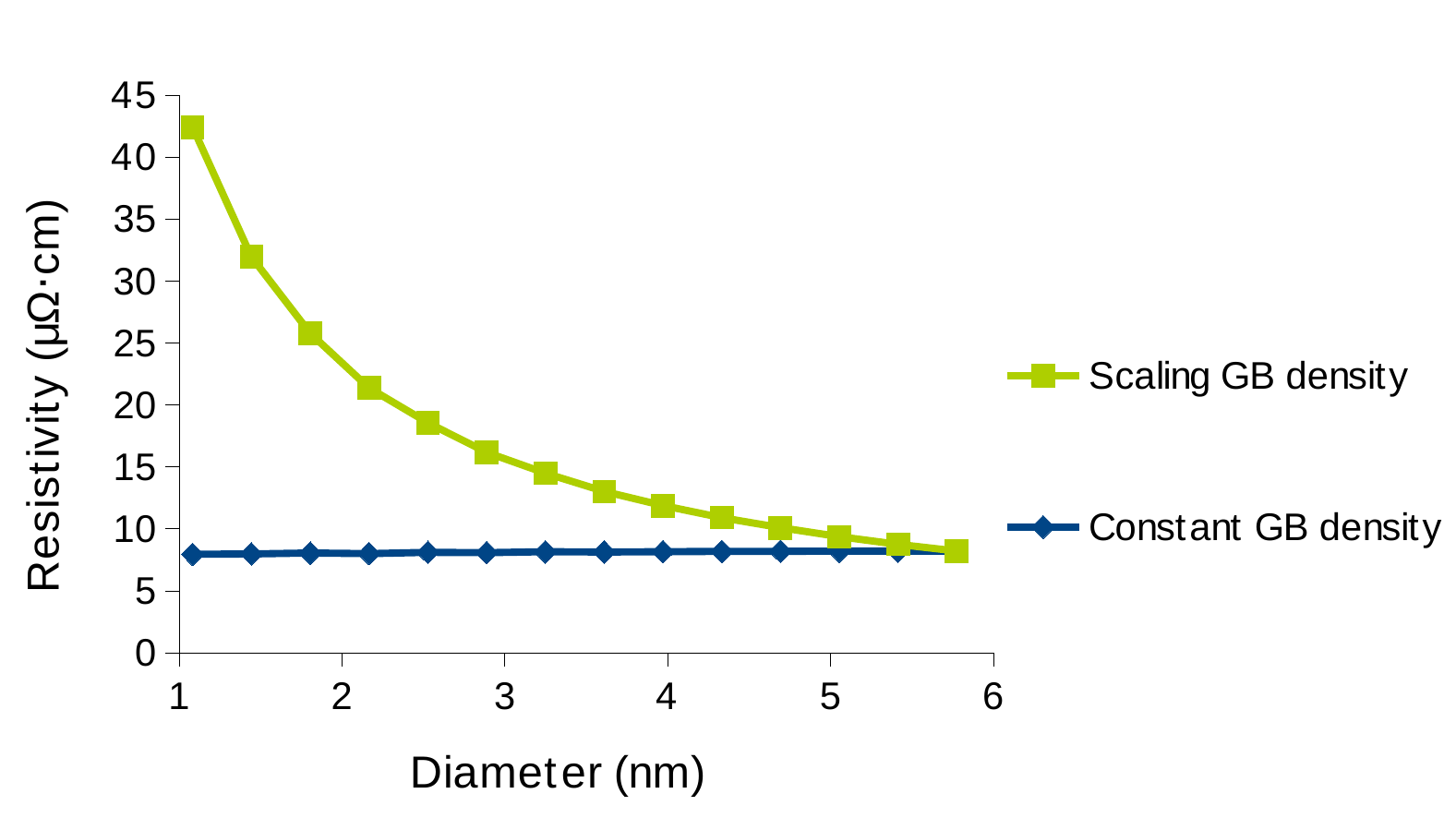}}
\caption{\label{resistivityScalingScenarios} The resistivity is shown for different diameter, copper nanowires with different types of surface roughness and grain-boundary properties. (a) The surface properties for the 16 by 16 $a_\textnormal{Cu}$ ($D \approx 5.8$ nm) wire are given by $\Delta = 0.1 D$, $\Lambda = 10 \textnormal{nm} \approx 1.75 D$ in all three cases. For the \textit{constant roughness} curve, the numerical values of $\Delta$ and $\Lambda$ are kept constant when simulating smaller nanowires. For \textit{scaling roughness size}, the roughness standard deviation is scaling as $\Delta = 0.1 D$, while $\Delta = 0.1 D$ and $\Lambda = 1.75 D$ are taken for the \textit{scaling roughness size and correlation} plot points. (b) For the two cases, the 16 by 16 $a_\textnormal{Cu}$ wire has an average intermediate distance of grain-boundaries equal to $D$, a standard deviation of $D/2$ potential barrier determined by $U_\textnormal{GB} = 1.5$eV, $L_\textnormal{GB} = a_\textnormal{Cu}$. For \textit{Scaling GB density}, the intermediate distance and standard deviation scale, whereas for \textit{Constant GB density} the numerical values of the 16 by 16 wire are taken for every diameter.}
\end{figure}

The resistivity, defined as the inverse of the conductivity, is different for the different methods (exact or approximate) of obtaining the relaxation times. The correct result, using Eq.~(\ref{exactRT}), is compared with results obtained with the approximations discussed in section \ref{sectionBTERTA} for both grain-boundaries and surface roughness in Fig.~\ref{resistivityScaling}. The resistivity is shown for Cu nanowires with diameters ranging from approximately 1 up to 6 nm. It is plotted for a discrete set of diameters, namely the integer multiples of the atomic lattice constant. Larger diameters above 6 nm can also be reached with our method, but the required computation time rapidly increases,\footnote{The computation time grows like the number of sub-bands squared, because the matrix elements have to be calculated separately for every pair of states. This results in calculation time growing more or less like $D^4$.} most of the CPU time being spent to calculating the surface roughness matrix elements.

Figure~\ref{resistivityScaling} (a) shows that neglecting the incoming scattering largely overestimates the resistivity. Assuming equal relaxation times gives a much better result, which, however, deviates substantially from the correct resistivity. This is plausible because incoming scattering of states with similar transport momentum $k_z$ can have a stabilizing effect on a particular current-carrying state. This effect is included in the exact solution as well as in the approximated solution relying on equal relaxation times, both having the same order of magnitude resistivity approximately a factor of 10 smaller than the other one. The resistivity is underestimated by the equal RT assumption because it overestimates the incoming scattering for the states carrying most of the current, thus enhancing their conductivity contribution. The underestimation of the resistivity may be compensated by the appearance of ``negative'' relaxation times, sometimes even leading to a much higher resistivity result. This effect can be seen in the deviations of the \textit{Equal RT} curve around 2 nm.

For grain-boundaries, in Fig.~\ref{resistivityScaling} (b), the approximations work a lot better. In this case neglecting the incoming scattering underestimates the resistivity, whereas equal relaxation times lead to the exact result, because there is only Umklapp scattering. The fully reflected states comply automatically with the equal RT assumption automatically because of symmetry. Scattering occurring only between states with the same sub-band indices is an artifact of the simple grain-boundary model and is not expected to emerge in more general grain-boundary modeling. Strangely enough, neglecting the incoming scattering underestimates the resistivity. Incoming scattering  of the opposite $k_z$ state equilibrium distribution is automatically included in the relaxation time, which is an overestimate in this case, explaining a lower resistivity.

In Fig.~\ref{resistivityScalingScenarios} different scenarios for surface roughness and grain-boundaries are considered. Remarkably, if the grain-boundary standard deviation is taken to be proportional to the diameter, the resistivity does not increase whether the correlation length scales with the diameter or not. Only when the standard deviation is constant for every diameter do we observe an increase. The values of resistivity are also quite small for all the data points, while the roughness standard deviation for every wire boundary of each data point is at least 10$\%$ of the wire diameter. There are also bumps in the resistivity plot showing up for all the roughness properties at certain diameters, independent of the surface roughness properties. This effect is due to sub-band quantization and it should be possible to confirm it with high-precision measurements, if the grain-boundaries were only modestly contributing. This appears to be not the case, as can be seen in Fig.~\ref{resistivityScalingScenarios} (b). Note that the resistivity values are proportional to $U_\textnormal{GB}$, the grain-boundary potential barrier. It is therefore very important to assign a realistic value to it. To this end, we have estimated its value using data in the 15-100 nm regime. \cite{waser2012nanoelectronics}\footnote{The bulk resistivity has been subtracted from the data points and 50$\%$ of the scaling part has been attributed to grain-boundaries.}

Unlike the case of FS and MS, no strict power-law relation between diameter and resistivity emerges. Nevertheless, we can still calculate an average scaling exponent by fitting the best power-law: $\rho \propto D^{\alpha}$ ($\alpha = -1$ for FS and MS models). The scaling exponents for all the simulated scenarios shown in Fig.~\ref{resistivityScaling} are given in Table \ref{scalingExponents}.
\begin{table}[h]
\centering
\caption{\label{scalingExponents} Approximate resistivity scaling exponents $\alpha$ ($\rho \propto D^\alpha$) are given for different surface roughness and grain-boundary characteristics.}
\begin{tabular}{lcc}
\hline
\hline
Scattering mechanism & Properties & $\alpha$ \\
\hline
Surface roughness & $\Delta \propto D$, $\Lambda =$ constant & 1.3 \\
 & $\Delta = $ const., $\Lambda = $ const. & -0.8 \\
 & $\Delta \propto D$, $\Lambda \propto D$ & 0.1 \\
Grain-boundaries & $L_z/N \propto D$ & -1 \\
 & $L_z/N = $ const. & 0 \\
\hline
\hline
\end{tabular}
\end{table}

\section{Conclusion}
We have developed a method to solve the scattering relaxation times exactly for electrons moving through a metallic nanowire. The method relies on an effective mass description of the electrons and the Boltzmann transport equation in the linear response regime, whereas the collision terms are determined by Fermi's golden rule. Our approach amounts to solving a system of coupled equations, providing the electron relaxation times for every individual state. Though being presented for zero temperature, the procedure can be generalized to non-zero temperatures by including more states to the system of equations.

The framework has been applied to retrieve the relaxation times for surface roughness and grain-boundary scattering in a metal nanowire with square cross-section. For both scattering mechanisms the relaxation times depend strongly on the considered sub-band state, the highest lifetime being reached for states with highest transport momentum, an effect that is pronounced for surface roughness scattering. For the grain-boundary model, the relaxation times are uniquely determined by the value of $k_z$ through a linear relation, while for surface roughness some spread for constant $k_z$ is observed and the relation is less clear.

Realistic surface roughness and grain-boundary properties show that the resistivity is largely dominated by grain-boundaries in nanowires with sides between 1 and 6 nm. The order of magnitude of the relaxation times can differ by a factor of 10-100 and fast traveling states are better protected against surface roughness scattering. When it comes to resistivity scaling in terms of the wire diameter, substantial differences with FS and MS scaling for large diameters are observed, both for surface roughness scattering and grain-boundary scattering. If the density of grain-boundaries increases inversely proportional to the diameter, the scaling law persists ($\alpha = -1$). However, a constant density would stop the scaling ($\alpha = 0$), while, if grain-boundaries can be avoided, a wide range of approximate scaling exponents related to surface roughness scattering ($-1 \le \alpha \le 1.5$ for surface roughness characteristics considered here) could be observed.

\begin{acknowledgments}
We would like to thank Christian Maes, Maarten Thewissen and Maarten Vandeput for many useful discussions, and Ruben Monten and Filip Sevenants for all the assistance with computer-related issues.
\end{acknowledgments}

\bibliography{paper}{}

\end{document}